\documentclass[12pt]{iopart}

\newcommand{\indexit}[1]{\mbox{\tiny #1}}
\usepackage{iopams}
\usepackage{graphicx}
\usepackage{multirow}
\usepackage{color}

\begin{document}

\title[Dynamical transitions and quantum quenches in mean-field models]{Dynamical transitions and quantum quenches in mean-field models}

\author{Bruno Sciolla and Giulio Biroli}

\address{Institut de Physique Th\'eorique, CEA/DSM/IPhT-CNRS/URA 2306 CEA-Saclay,
F-91191 Gif-sur-Yvette, France}
\ead{bruno.sciolla@cea.fr}
\begin{abstract}
We develop a generic method to compute the dynamics induced by quenches in completely connected quantum 
systems. These models are expected to provide a mean-field description at least of the short time dynamics of finite dimensional system.
We apply our method to the Bose-Hubbard model, to a generalized Jaynes-Cummings model, and to the Ising model in a transverse field. We find that the quantum evolution can be mapped onto a classical effective dynamics, which involves only a few intensive observables. For some special parameters of the quench, peculiar dynamical transitions occur. They result from singularities of the classical effective dynamics and are reminiscent of the transition recently found in the fermionic Hubbard model. Finally, we discuss the generality of our results and possible extensions.
\end{abstract}

\maketitle

\section{Introduction}
\label{intro}
Thanks to the fast experimental progress on cold atoms \cite{greiner,greiner2,greiner3}, where direct control over the parameters of the effective Hamiltonian is available, many theoretical questions about out of equilibrium dynamics of quantum systems have been raised and started to be addressed.  Among them, we cite the problem of thermalization for an isolated system (see e.g. \cite{rigol, biroli}) the existence of  long-lived out of equilibrium states (see e.g. \cite{weiss,kollath,roux,roux2}), dynamical transitions out of equilibrium \cite{werner,schiro,here,calabrese}, etc.

In this work we focus on the off-equilibrium dynamics induced by quantum quenches, {\it i.e.} the evolution of an isolated quantum system after a sudden change of a parameter of the Hamiltonian. 
This problem has been intensively studied  these last years. It provides an useful idealization of the protocols 
followed in experiments, where the change of parameters takes instead place at finite rates. 
The literature on quantum quenches is already quite broad, see for example the reviews \cite{review1,review2,review3}. 
One dimensional systems have been intensively studied by exact analytical and numerical methods. 
Results for higher dimensional systems are instead scarcer; the previous methods cannot be applied and one has to 
resort to approximations of some kind. The ones that have been more used are the Bogoliubov method \cite{Zurek}, 
path integral saddle-point expansions \cite{altman,kennett}, the large-$N$ limit \cite{Fischer} and mean-field (or large dimension) approximations \cite{mds,werner,schiro,here,snoek}. 
Interestingly, mean-field approaches revealed that quantum quenches may lead to out of equilibrium dynamical transitions. This was first found in the analysis of the Fermionic Hubbard model by time dependent Dynamical Mean Field Theory (t-DMFT) in \cite{werner}. Later, a confirmation and explanation of this result was found by 
using a Gutzwiller time-dependent Ansatz in \cite{schiro}. In \cite{here} we solved the Bose Hubbard model
 on a completely connected lattice and found a very similar dynamical transition at integer fillings. This was also noticed in the hard-core Bose-Hubbard model in a superlattice potential \cite{wolf}. Finally, clues to argue that these dynamical transitions are generic even beyond mean-field
 were recently presented in \cite{calabrese} by mapping the problem to the one of classical phase transitions in films.

In this article we first describe a generic approach to solve the dynamics of quenches in completely connected quantum models. Then, we apply our method to the Bose-Hubbard model, the generalized Jaynes-Cummings model and the Ising model in a transverse field. We find that the dynamical transition, discovered for the Hubbard model, is a systematic dynamical effect present in completely connected systems characterized by a quantum phase transition in equilibrium. A shorter version of this work, that focused only on the Bose-Hubbard model, appeared in \cite{here}. 

\section{Summary of the method and results}

In this article, we focus on the out of equilibrium dynamics in several completely connected models:
the Bose Hubbard model (BHM), the transverse field Ising model (IM) and the Jaynes-Cumming model (JCM). 
The reason for focusing on systems defined on completely connected graphs is that this allows one to study a well defined model and---at the same time---to obtain an approximate solution for finite dimensional lattices. Indeed, 
such models are related to several approximations used in the literature. Among the most known ones, we cite the limit of infinite dimensions for models defined on a hyper-cubic $d$-dimensional lattice\footnote{This consists in a technique similar to the ones developed in \cite{Georges}.} and the Gutzwiller 
Ansatz, a widespread method to obtain mean field equations. Actually, in the case of completely connected models, the Gutzwiller Ansatz can be shown to be exact by using a Hubbard-Stratonovich transformation. 

We shall focus on out of equilibrium dynamics induced by a quantum quench. This procedure is defined as follows. Let $\hat{H}(\lambda)$ be the quantum Hamiltonian, which depends on a coupling $\lambda$. The system is prepared in its ground state $| \psi (t<0)\rangle = |\textrm{GS}(\lambda_i) \rangle$ at some coupling $\lambda_i$.
At $t\geq0$, the system is driven out of equilibrium in a controlled fashion, tuning the coupling $\lambda(t)$ in time according to a predefined procedure. 
Except for quasistatic procedures, often called ``adiabatic'', the wave function becomes different from the ground state $| \psi (t>0)\rangle \neq |\textrm{GS}(\lambda(t)) \rangle$.
We call \emph{sudden quench} the procedure in which the coupling is suddenly switched to a final value $\lambda_f$: $\lambda(t) = (\lambda_f -\lambda_i) \; \theta(t) + \lambda_i$.

As we shall show, the quantum dynamics in a completely connected system can be solved by mapping the unitary evolution onto an effective model undergoing Newtonian dynamics. This is a drastic simplification which is possible thanks to the symmetry of the completely connected Hamiltonian under any permutation of sites. 
We restrict our analysis 
to cases where $|\psi(t=0) \rangle$  is the ground state at some coupling $\lambda_i$. Thus, $|\psi(t=0) \rangle$ is also symmetric under permutation of sites and since both the initial state and the Hamiltonian are symmetric, the generic unitary evolution takes place in the sector of symmetric states only. In this subspace, the states are parametrized using a few local macroscopic observables. The unitary evolution can then be written as a Schr\"odinger equation in the symmetric space, which involve an effective $\hbar = V^{-1}$, where $V$ is the number of sites of the system. Thanks to this property, at the thermodynamic limit, the entire dynamics of the system can be encoded in the one of few macroscopic variables, which undergo an effective classical Hamiltonian evolution.

We first apply this approach to the Bose-Hubbard model, our main interest because of its applicability to cold atom experiments. In this case, it is the onsite repulsion $U$ that plays the role of the coupling $\lambda$.  
In a first stage we make the additional assumption that the number of bosons per site is less or equal to $n_b^{max}=2$.
The average superfluid order after the quench is a non-monotonous function of $|U_f-U_i|$. 
Actually, it decays logarithmically to zero at some special values of $U_i$ and $U_f$.
For these special quenches, the superfluid order relaxes exponentially to zero. We call this peculiar feature a dynamical transition. Surprisingly, the microcanonical equilibrium characterized by same energy, towards which the system would relax on large times if it were able to thermalize\footnote{Thermalization is very likely to occur in finite dimensions for the Bose-Hubbard model, because the system is not integrable. Instead for completely connected models we do not expect thermalization to the Gibbs ensemble, as we shall discuss later. This implies that in the large dimension limit the timescale for equilibration diverges as a function of $d$.}, has non-zero superfluid order.
Thus, this transition is a purely \emph{dynamical} phenomenon, which has nothing to do with the fact that the superfluid order vanishes in the high temperature phase. 
We extend our analysis to $n_b^{max} \geq 3$. In this case the effective dynamics involves $n_b^{max}-1$ degrees of freedom, and the classical trajectories can be either regular or chaotic. Apart from this difference, the dynamical transition is found to be qualitatively unchanged and the quantitative differences with $n_b^{max}=2$ are small.

Finally, we focus on two other systems, for which some out of equilibrium properties have already been studied, a generalized Jaynes-Cummings model \cite{Gurarie}, and the Ising model in a transverse field \cite{mds}. We apply 
our formalism to both systems and show that also in these cases there is a dynamical transition for sudden quenches. 

This paper is organized as follows: In section \ref{models}, we briefly describe the three considered models. In Section \ref{general}, we describe how the effective classical dynamics can be derived for an arbitrary Hamiltonian. Then in section \ref{special}, we derive the classical dynamics for the Bose-Hubbard model with truncation $n_b^{max}=2$, and analyze quenches and the dynamical transition in section \ref{quenches}.
The general case $n_b^{max}\geq 3$ is considered in section \ref{more}. The effective dynamics and quench properties of the generalized Jaynes-Cummings model are described in section \ref{Jaynes}, and in section \ref{Ising} for the Ising model. Section \ref{discussion} and \ref{conclusion} contain discussions and possible extensions of our work.
\ref{Gutzwiller} is devoted to prove that the same effective dynamics is recovered with a Gutzwiller Ansatz wave function, and in \ref{wkb_expansion} the WKB eigenstates of the completely connected model are discussed.

\section{Definition of the models}
\label{models}
\subsection{Bose-Hubbard model and the Mott-Superfluid transition}

The Bose-Hubbard lattice model has regained a lot of interest recently since it can be realized and studied 
in experiments on cold atoms \cite{Jaksch}. The Hamiltonian of its completely connected version, suited to describe the limit $d \rightarrow \infty$ of the lattice model, reads
\begin{equation}
\label{hamiltonian}
 H = \frac{U}{2}\sum_i n_i(n_i-1) -\frac{J}{V} \sum_{i \neq j} b^\dagger_j b_i
\end{equation}
where $b^\dagger_i$, $b_i$ are the bosonic creation and annihilation operators, satisfying $[b_i, b^\dagger_j] = \delta_{ij}$, and $n_i = b^\dagger_i b_i$. The first term is an on-site repulsion between bosons, the second is a tunneling term, of amplitude $J$, and rescaled by the volume $V$ (number of sites). This rescaling is needed to obtain a well-defined thermodynamic limit $V \rightarrow \infty$. The limit of infinite dimensions that is equivalent to this model is obtained
by taking a coupling $\frac{J}{2d}$ for a BHM defined on a $d$-dimensional hyper-cubic lattice.

This model undergoes a quantum phase transition at \emph{commensurate} fillings: $\langle \hat{n}_i \rangle = n$ with $n$ integer. See \cite{Fisher} for a detailed discussion of the phase diagram. Notice that since we consider 
the dynamical behavior, the number of particles is fixed because it corresponds to a quantity 
conserved by the dynamics. In consequence we shall discuss the phase diagram in terms of the density and not the 
chemical potential, as it is instead done usually. 
Let us focus first on cases where the number of particles per site is an integer. To grasp why there is a quantum phase transition, it is helpful to compare the ground state in the two limits $U \rightarrow \infty$ (or $J=0$) and $U\rightarrow 0$.
In the first case, the ground state is diagonal in the occupation number $|\psi \rangle = \otimes_i | n_i=n \rangle $. It remains
the same for all $U<U_c$, and is called a Mott insulator. It is incompressible because adding or removing a boson requires an energy of the order of $U$.
In the second case, $U=0$, the ground state is akin to\footnote{Formally, this state is the true ground state only in the grand canonical ensemble. However, the canonical and grand canonical ensemble are equivalent up to $1/V$ corrections.} a product of coherent states $|\psi \rangle = C \prod_i e^{-\alpha b^\dagger_i} |0\rangle$.
For $U=0$, and also moderate values of $U$, the ground state is superfluid and compressible. The usual order parameter $\langle \hat{b}_i\rangle$ is zero because the density is fixed, hence the appropriate parameter is the off-diagonal long range order measured by $|\Psi_0|^2 = \lim_{d_{ij} \rightarrow \infty} \langle \hat{b}^\dagger_j \hat{b}_i\rangle$. In the completely connected model, all distances $d_{ij}$ between two different sites are equal to $1$. This is the 
largest distance in the problem. In consequence one can define\footnote{In equilibrium, it is easy to check that the definition below gives back the usual one used in the grand-canonical ensemble, $|\Psi_0|^2=|\langle b \rangle|^2$}: 
\begin{equation}
\label{superfluid_order}
|\Psi_0|^2 = \langle \hat{b}^\dagger_j \hat{b}_i \rangle\qquad i\neq j
\end{equation}
Because the system is completely connected, phonons are absent and the spectrum always has a gap. Beyond a critical coupling $U_c$, the order parameter $|\Psi_0|^2$ vanishes and the ground state becomes a Mott insulator. This is the Mott-superfluid quantum phase transition. For non-integer fillings, the system is instead always superfluid and compressible as it can be understood considering perturbation around the $U\rightarrow \infty$ limit.
At non zero temperature, there is a second order phase transition from the superfluid phase to a Bose gas at a critical temperature $T_c(U)$.

\subsection{Generalized Jaynes-Cummings model and superradiance transition}

The second model that we consider is the generalized Jaynes-Cummings model studied in \cite{Gurarie}, which describes $N$ distinguishable two-level systems in interaction with a single quantized electromagnetic mode.
It is useful in various contexts, and has been suggested as a qualitative description of the BEC/BCS crossover in fermionic condensates, of molecular magnetism and of the formation of dimers by pairing of two bosons. Recently, new experiments in the setting of cavity quantum electrodynamics \cite{Baumann} are realizations of the Dicke Hamiltonian \cite{Keeling}, which can be mapped onto the Jaynes-Cummings Hamiltonian if the ``rotating wave approximation'' is made.
The dynamics of sweeps, starting from an empty bosonic mode, was studied in \cite{Gurarie} using quasiclassical approximations and the truncated Wigner approximation. Here, we describe the dynamics of sudden quenches from the broken symmetry phase.

The model is defined as follows. There is one bosonic $[\hat{b}, \hat{b}^\dagger]= 1$ degree of freedom---the electromagnetic mode---and one SU(2) spin $S=N/2$. The spin degree of freedom keeps track of the number of excited two-level systems, $n_{\mbox{exc}} = S^z+N/2$. With $\hat{n} = \hat{b}^\dagger \hat{b}$, the Hamiltonian is
\begin{equation}
\label{Jaynes_Hamiltonian_0}
\hat{H} = \omega_0 \hat{S}^z + \omega \hat{n} + \frac{g}{\sqrt{N}}(\hat{b}^{\dagger}\hat{S}^- + \hat{b}\hat{S}^+)
\end{equation}
It includes a potential energy for bosons ($\omega$), for the spin ($\omega_0$), and a coupling $g$ between the two.
Each time a boson is lost, a two-level system is excited, and reciprocally: the quantity $\hat{Q}= \hat{S}^z + \hat{b}^{\dagger}\hat{b}$ is conserved by the dynamics. It is thus possible to define $2\lambda = \omega_0 - \omega$ such that up to a constant term one finds:
\begin{equation}
\label{Jaynes_Hamiltonian}
\hat{H} = -2\lambda \hat{n} + \frac{g}{\sqrt{N}}(\hat{b}^{\dagger}\hat{S}^- + \hat{b}\hat{S}^+)
\end{equation}
In the following we take $g=1$ and consider $\lambda$ in units of $g$, and the time in units of $\hbar/g$.  We consider the regime $Q \geq N/2$ (which can be reduced to $Q = N/2$), for which there is a quantum phase transition.  At $T=0$, the system is in the \emph{normal} ground state $\langle \hat{n} \rangle= 0$ if $\lambda<\lambda_c=-1$ and is in the \emph{super-radiant} ground state for $\lambda>\lambda_c$ where $\langle \hat{n} \rangle \neq 0$.
This is the super-radiance quantum phase transition. This transition is the consequence of the competition between the 
two terms in the Hamiltonian: the first one favors removing bosons (if $\lambda<0$), whereas the second 
plays the role of a kinetic energy and it can lower the energy if the bosons modes are filled, see section \ref{Jaynes}.
As for the Bose-Hubbard model, there is a finite temperature phase transition at $T_c(\lambda)$, above which the super-radiant phase disappears. 
Notice that this phase transition is usually studied in the canonical/microcanonical ensemble, whereas here 
it is more natural to focus on given value of $Q$ since this quantity conserved by the dynamics. 

\subsection{Ising model in a transverse field and the ferromagnetic transition}
The Ising model in a transverse field is a paradigm of quantum phase transitions \cite{sachdevbook}. The Hamiltonian of its completely connected version reads
\begin{equation}
\hat{H} = - \frac{J}{2N} \sum_{ij} \hat{S}^z_i \hat{S}^z_j - \Gamma \sum_i \hat{S}^x_i
\end{equation}
where $J$ is the ferromagnetic coupling and $\Gamma$ the transverse field. For simplicity, we set $J=1$, measure $\Gamma$ in units of $J$ and time in units of $\hbar/J$. The Hamiltonian can be written using a single spin $\vec{\hat{S}} = \sum_i \vec{\hat{S}}_i$ and reads
\begin{equation}
\hat{H} = -\frac{1}{2N} (S^z)^2 - \Gamma S^x
\end{equation}
The large $N$ limit corresponds to the large spin $S=N/2$ limit, which is also the classical limit. The ground state is found, minimizing the corresponding classical Hamiltonian\footnote{The kinetic term of the classical limit is nontrivial to obtain, but we do not need it for the moment.} $\vec{\hat{S}}\rightarrow \vec{S} = S\{\sin\theta \cos\phi, \sin\theta \sin\phi, \cos\theta \}$ \cite{mds}.
The minimum is at $\phi = 0$, and for $\Gamma < 1/2$, $\sin \theta = 2 \Gamma$, which is a \emph{ferromagnetic} ground state, in which the spins align toward the $z$-axis.
For $\Gamma > 1/2$, $\theta = \pi$ and the ground state is called \emph{quantum paramagnet}, because it is unoriented in the $z$-basis $| \pm \rangle_i$, $|\psi\rangle = \frac{1}{\sqrt{2}^V} \otimes_i \left( | +\rangle_i +| -\rangle_i \right)$.
The quantum phase transition from z-ferromagnet to quantum paramagnet takes place at $\Gamma_c = 1/2$.
Starting from the ferromagnetic ground state and increasing the temperature the system undergoes a second-order phase transition at a finite temperature $T_c(\Gamma)$.

\section{Quantum quenches in completely connected models: mapping to an effective classical dynamics}
\label{general}
\subsection{Site permutation symmetry and classical Hamiltonian dynamics}

In the following, we show how the dynamics induced by quantum quenches, in arbitrary completely connected models, can be mapped onto an effective classical Hamiltonian dynamics.
As already stated, any model defined on a completely connected lattice has a Hamiltonian which is symmetric under any permutation of sites. Thus, one expects that  the ground state of the model is {\sl also} site permutation symmetric\footnote{This symmetry may be spontaneously broken in systems with attractive interactions, which we do not consider here.}.
We shall study quenches starting from the ground state of the system. Since the site permutation symmetry is conserved by the unitary evolution, the dynamics only takes place into the subspace of symmetric wave functions. This is a drastic simplification, because symmetric states can be described using a few variables only.

To clarify this point in a concrete but simple case, let us see what these variables are for the Bose-Hubbard model, with the additional constraint that there are $n_b^{max} = 2$ or less bosons per site. First, remark that given a particular Fock state $|\{n_i\} \rangle$, the linear combination of every possible site permuted Fock state is a site permutation symmetric state $|S(\{n_i\}) \rangle \sim \sum_P |\{n_{P(i)}\} \rangle$.
This new state is completely characterized by the fraction $x_0$, $x_1$, $x_2$ of sites with $0$, $1$ and $2$ bosons respectively, and we call it $|x_0,x_1,x_2\rangle = | x\rangle$ where $x$ is a shorthand for all variables. Since the Fock states form a basis of generic states, the $\{ |x\rangle \}$ states form a basis of the symmetric sector.
In order to express the Schr\"odinger evolution in this basis, we have to compute the transitions elements $\langle x' |\hat{H}| x \rangle$. Typically, only few transitions $W_{m}(x) = - \frac{1}{V}\langle x + m/V |\hat{H}| x \rangle$ are allowed\footnote{The minus sign in this definition is for later convenience.}, with $m = \{m_0,m_1,\ldots\}$ a vector of integers.
For example in the Bose-Hubbard model, the matrix elements of the Hamiltonian connect states that differs by one boson jump. Thus $x_0$, $x_1$ and $x_2$ can only differ by $1/V$ or $2/V$.
For example, after a jump $|2_i \ldots 0_j \ldots\rangle \rightarrow |1_i \ldots 1_j \ldots\rangle$, $x'_1 = x_1 +2/V$, $x'_0 = x_0 -1/V$ and $x'_2 = x_2 -1/V$. This transition is labeled $m = \{m_0 = -1, m_1 = 2, m_2 = -1 \}$. Moreover, the reverse move is a transition characterized by $-m$, with same amplitude $W_{-m}(x) = W_{m}(x)$ at dominant order in $V$.

%

Thanks to the ``locality'' of transition elements between symmetric states, the Schr\"odinger equation in the symmetric basis $|\psi \rangle = \sum_{x} \psi_{x}(t) |x\rangle$ takes a simple form. The Schr\"odinger equation projected on $\langle x |$ reads $\langle x |i \partial_t|\psi\rangle =\langle x |H| \psi\rangle$. Denoting the diagonal transition element $D(x) = \frac{1}{V}\langle x|H|x\rangle$, we get
\begin{equation}
\label{gen_schrod}
\begin{array}{ll}
\displaystyle i \partial_t \psi_{x} & \displaystyle = V D(x) \psi_{x} - V \sum_{m} W_{m}(x) \Bigl( \psi_{x+m}+\psi_{x-m} \Bigr) \\
\displaystyle & \displaystyle = V \Bigl( D(x) - 2 \sum_{m} W_{m}(x) \cosh(m_i \partial_{x_i} /V) \Bigr)  \psi_{x}
\end{array}
\end{equation}
where we made use of $\psi_{x + m/V} = \exp(m_i \partial_{x_i} /V)\psi_{x}$ with an implicit summation over the index $i$. Strikingly, the Schr\"odinger equation \eref{gen_schrod} involves an effective $\hbar = 1/V$, thus the regime of interest $V \rightarrow \infty$ corresponds to the \emph{classical regime}. Furthermore, for the ground state, the initial wave function is a narrow wave packet of width $1/\sqrt{V}$ (this general property is discussed in more detail for the Bose-Hubbard model in section \ref{classical_evolution_of_packet}). Thus, by the Heisenberg uncertainty principle, we expect that momentum fluctuations are proportional to $\sqrt \hbar \propto \sqrt{1/V}$. Therefore the evolution of the wave-packet can be fully described in the thermodynamic limit  
by its average (or center) $x(t) = \langle \hat{x} \rangle$ and average momentum $p(t) = \langle \hat{p} \rangle$. Both quantities evolve following a classical Hamiltonian dynamics obtained from the quantum one by
replacing  $\hat{p} = \frac{i}{V}\partial_x \rightarrow p(t)$ and $\hat x \rightarrow x(t)$:
\begin{equation}
\begin{array}{ll}
\displaystyle \frac{i}{V} \partial_t \psi_x  &= \Bigl( D_x - 2 W_x \cosh(2 \partial_x/V) \Bigr)  \psi_x\\
 &= \Bigl( D_x - 2 \sum_{m} W_{m}(x) \cos(m_i \hat{p}_i) \Bigr)  \psi_x = \hat{H}\psi_x \\
\end{array}
\end{equation}
\begin{equation}
\label{classical_Hamiltonian_generic}
H[x,p] = D(x) - 2 \sum_{m} W_{m}(x) \cos(m_i p_i)
\end{equation}
The classical Hamiltonian evolution of the variables $\dot{x}_i(t) = \partial H/\partial p_i$ and $\dot{p}_i(t) = -\partial H/\partial x_i$ yields the evolution of the wave packet after the quantum quench, and give access to all observables as a function of time.
A more careful analysis is performed in section \ref{rigorous} and fully supports this picture.

As a conclusion, the analysis of quench dynamics in connected models is tractable, and takes the form of an effective classical Hamiltonian evolution.
The initial condition is provided by the ground state values of $x_i,p_i$ at the initial coupling $U_i$. The sudden quantum quench dynamics is then described by the classical dynamics of $x(t),p(t)$ induced by the effective Hamiltonian $H(U_f)$, with initial conditions $x_i,p_i$.

\section{Effective classical Hamiltonian equations in the Bose-Hubbard model with truncation $n_b^{max}= 2$}
\label{special}

In the following, we study the Bose-Hubbard model with two or less bosons per site. We derive explicitly 
the effective classical Hamiltonian in section \ref{Bose-Hubbard_Hamiltonian}. 
The nature of the ground state, in particular the peaking of the wave packet when $V\rightarrow \infty$
and the quantum phase transition, are discussed in  section \ref{classical_evolution_of_packet}.
A more thorough derivation of the effective Hamiltonian dynamics is presented  in  section \ref{rigorous}.
Finally, we conclude by describing the phase space of effective trajectories in section \ref{types}. 

\subsection{Schr\"odinger equation in the site permutation symmetric basis and effective Hamiltonian}
\label{Bose-Hubbard_Hamiltonian}
In this section, the method sketched above is applied to the Bose-Hubbard model with truncation $n_b \le n_b^{max}=2$.
This restriction is well-suited to derive analytical expressions, and preserves the main features of the model, such as the Mott insulator to superfluid quantum phase transition.
The case of a larger number of bosons per site is considered in section \ref{more}. The results remain qualitatively the same.

The symmetric states are written under the form $| x_0, x_1, x_2 \rangle$. Since the total number of sites $V= V (x_0+x_1+x_2)$ is fixed, and the overall density of bosons $n = N/V = x_1+2x_2$ is conserved by the dynamics, the symmetric states can be labeled by one variable only.
Choosing $x_1$ as such a variable, the symmetric normalized wave functions are denoted $|x_1 \rangle = (N_0!N_1!N_2!/V!)^{1/2} \sum_{\{n_i\}}' |n_1,n_2,...,n_V \rangle $, where $N_i = V x_i$ and $\sum'$ means the sum over Fock states of fixed fraction $x_1$.
In the following for simplicity of notation we drop the subindex one and use $x$ instead of $x_1$. Note that contrary to the previous paragraph, $x$ is now just a number and not a vector. \\
To compute the transition rates, we proceed as follows.
The repulsion term $1/2\sum_i n_i(n_i-1)$ is diagonal in the $|x \rangle$ basis. The tunneling term $-1/V \sum_{i \neq j} b^\dagger_j b_i$ allows transitions from $|N_0,N_1,N_2 \rangle$ to three states, $|N_0-1,N_1+2,N_2-1 \rangle$, $|N_0+1,N_1-2,N_2+1 \rangle$ and $|N_0,N_1,N_2 \rangle$.
Using the previous notations, the only possible transitions correspond to $m = 2$, $m = -2$, $m=0$ representing respectively $\langle x+2/V|\hat{H}|x\rangle$, $\langle x-2/V|\hat{H}|x\rangle$ and $\langle x|\hat{H}|x \rangle$. The first amplitude reads
\begin{equation}
\begin{array}{l}
\langle x + 2/V |\hat{H}|x \rangle = \\
\displaystyle \left(\frac{(N_0-1)!(N_1+2)!(N_2-1)!}{V!}\frac{N_0!N_1!N_2!}{V!}\right)^{1/2} \sum_{\{n'_i\}}' \langle \{n'_i\} |\hat{H}   \sum_{\{n_i\}}  ' | \{n_i\} \rangle \\
\end{array}
\end{equation}

There are $V!/(N_0!N_1!N_2!)$ factors on the ket side. For each of these factors, there are $N_0 N_2$ transition to a state with $N_1' = N_1+2$. Therefore using that $b_i^\dagger b_j |0_i\rangle|2_j\rangle = \sqrt{2} |1_i\rangle|1_j\rangle$, and after a partial cancellation of the normalization factors, we obtain:
\begin{equation}
\langle x + 2/V |\hat{H}|x\rangle = \frac{-\sqrt{2}}{V} [N_0 N_2(N_1+1)(N_1+2)]^{1/2}
\end{equation}
In the following, only dominant contributions of order $V$ are kept, the sub-leading ones are dropped since they
do not matter for dynamics taking place on times not diverging with the system size. Rewriting $x_0$ and $x_2$ as functions of $x$, the transition rates are
\begin{equation}
\label{trans_elem}
\begin{array}{llll}
\displaystyle \langle x -2/V|\hat{H}|x \rangle & = & -V \, x [(2-x-n)(n-x)/2]^{1/2} & \equiv - V\, \hat W_x \\
\displaystyle \langle x |\hat{H}|x \rangle & = & V \, U(n-x)/2 - V \,  x(2+n-3x)/2 & \equiv V\, \hat D_x\\
\displaystyle \langle x +2/V|\hat{H}|x \rangle & = & \langle x -2/V|\hat{H}|x \rangle & \\
\end{array}
\end{equation}

The Schr\"odinger evolution on $\psi_x(t)$ in the site permutation symmetrical basis is $\langle x |i \partial_t|\psi\rangle =\langle x |H| \psi\rangle$
\begin{equation}
\begin{array}{ll}
\label{Schrodinger}
\displaystyle i \partial_t \psi_x  &=  V \, \hat D_x \psi_x - V \, \hat W_x \Bigl(\psi_{x+2/V} + \psi_{x-2/V} \Bigr) \\
 &= V \Bigl( \hat D_x - 2\hat W_x \cosh(2 \partial_x/V) \Bigr)  \psi_x \\
 &= \Bigl(\hat D_x - 2\hat W_x \cos(2\hat{p}) \Bigr)  \psi_x \,
\end{array}
\end{equation}
Therefore we have obtained that the Hamiltonian in the subspace of symmetric states reads 
\begin{equation}\label{hamiltoniann2}
\hat H=\hat D_x - 2\hat W_x \cos(2\hat{p})
\end{equation}
Note that the dimension of the Hilbert space has been reduced from $e^V$ to $V$ in this case. The case $n_b^{max} =2$ is especially convenient because the effective classical motion takes place in one dimension, therefore the equation of
motion are integrable and the evolution easy to understand and describe.

\subsection{Nature of the ground state and quantum phase transition}
\label{classical_evolution_of_packet}
As in section \ref{general}, the Schr\"odinger equation \eref{Schrodinger} involves an effective $\hbar = 1/V$, and its classical limit is obtained through $\hat{x} \rightarrow x(t)$ and $\hat{p}\rightarrow p(t)$.
\begin{equation}
\label{classical_Hamiltonian}
H[x,p] = D(x) - 2 W(x) \cos(2p)
\end{equation}

The possibility of describing the quantum dynamics after a quench in terms of classical dynamics relies on the fact that the initial wave function is initially of small width ($\sim 1/\sqrt{V}$). In order to check this, 
let's first expand the eigenvalue equation at lowest order in $1/V$:  \[E \psi_x = (\hat D_x - 2\hat W_x - \frac{4 \hat W_x }{V^2} \partial^2_x) \psi_x\]
Since the kinetic term has a factor $1/V^2$ in front, the ground state corresponds to the minimum of $ D_x - 2W_x$ and small quantum fluctuations around it. Indeed, around the minimum, $x_{\indexit{GS}}$, a quadratic expansion of $W_x$ and $D_x$ maps this problem onto a quantum harmonic oscillator with $m = W_{x_{\indexit{GS}}}/8$ and $\omega = \sqrt{\partial_x^2(D-2W)|_{x_{\indexit{GS}}}/m}$.
Thus, we find that the ground state wavefunction is centered around the absolute minimum $x_{\indexit{GS}}$ of the potential $D(x) - 2W(x)$ and
has a width $\sigma \sim \sqrt{\hbar /m \omega}$, which is of the order of $1/\sqrt{V}$. An appropriate expression for any eigenstate is provided by the WKB expansion is given in section \ref{wkb_expansion}.


The quantum phase transition taking place as a function of $U$ can be recovered within this formalism as we now show. We recall that for the Bose-Hubbard model, the quantum phase transition and the Mott phase are present at commensurate fillings only, so we have to focus on $n=1$ for $n_b^{max}=2$.
The ground state of the quantum Hamiltonian corresponds to the global minimum of the effective Hamiltonian \eref{classical_Hamiltonian}, given by $\frac{\partial H}{\partial x}=\frac{\partial H}{\partial p} = 0$.
These equations lead to $p = 0$ and $x$ equal to the value $x_{\indexit{GS}}$, which verifies $\frac{\partial(D(x) - 2W(x))}{\partial x} =0$. It is easy to check that 
\begin{equation}
x_{\indexit{GS}} = \left \{
\begin{array}{lll}
\displaystyle 1 & \textrm{ if } U \geq U_c, & \textrm{Mott insulator ground state}\\
\displaystyle \frac{U/U_c +1}{2} < 1 & \textrm{ if } U<U_c, & \textrm{Superfluid ground state}\\
\end{array} \right. 
\end{equation}
where $U_c = 3 + 2\sqrt{2} \simeq 5.82843$ (note that $x\le 1$). 
In the thermodynamic limit, the Mott insulator ground state is $|n_0 = 1, n_1 = 1 \dots \; \rangle = |x = 1\rangle$. Instead the superfluid ground state is $|x = x_0\rangle$, with $1/2<x_0<1$. Therefore, in the superfluid state, a fraction of sites  $x_2 = (1-x_{\indexit{GS}})/2>0$ contain two bosons.
$|\Psi_0|^2 = \frac{1}{V^2} \langle \sum_{ij} b^\dagger_i b_j \rangle$ is proportional to the intensive kinetic energy, and can be computed using the identity $|\Psi_0|^2 = -1/J (E - U/2 \langle \sum_i n_i(n_i-1) \rangle) = x_{\indexit{GS}}(1-x_{\indexit{GS}})U_c/2$.
It is the order parameter for the transition: it is positive in the superfluid phase and vanishes in the Mott insulating one.


\subsection{Effective classical evolution of wave packets}
\label{rigorous}
In the following we show in detail how the classical Hamiltonian dynamics emerges from the quantum evolution of symmetric states in the thermodynamic limit.
For simplicity we consider the situation of section \ref{special}, where $n_b^{max}\le 2$, but the argument is more 
general. The Schr\"odinger equation reads
\begin{equation}
\label{Schrodinger2}
\frac{1}{V} i \partial_t \psi_{x} = \,\Bigl( D_{x} -  \, 2W_{x} \cosh (2\partial _{x}/V)\Bigr)\psi_{x}
\end{equation}
A ground state wave function is a wave packet characterized by a small width of the order of $1/\sqrt{V}$ as shown in section \ref{classical_evolution_of_packet}. The width broadens after a long time, which diverges with $V$. On times that do not diverge in the thermodynamic limit, the ground state wave function and its subsequent evolution can be written as 
\begin{equation}
\label{psiexp}
\psi(x,t) = e^{-Vf(x,t)}
\end{equation}
 with $f(x,t) = g(x,t) - i \theta(x,t)$, $g$ and $\theta$ being respectively the envelope and the phase of the wave-packet. The envelope $g(x,t)$ has a maximum at $x(t)$, and expanding $g(x,t)$ around $x(t)$ shows that indeed this wave function describes a packet of width $1/\sqrt{V}$.

\begin{figure}
\centering
  \includegraphics[scale=0.8]{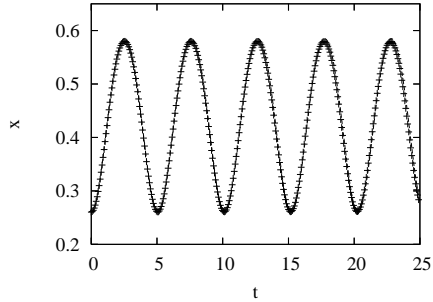}
\caption{The trajectory $x(t)$ for $U=3.33$, with initial conditions $x(0)=0.26$ of energy $E = 0.2$. The dots are obtained by numerical diagonalization of \eref{Schrodinger} for $V=5000$ with a sharp Gaussian initial condition. The line is the evolution according to the classical Hamiltonian. The two are equivalent on short times. Similar results hold for any trajectory and initial condition.}
\label{fig:comparison}
\end{figure}

After these preliminary considerations, we proceed and evaluate the evolution of the wave function, plugging the expression (\ref{psiexp}) of $\psi(x,t)$ into (\ref{Schrodinger2}):
\begin{equation}
-i \partial_t f(x,t) = D_x - 2 W_x \cosh(2\partial_x f) 
\end{equation}
For the amplitude and the phase, this leads to:
\begin{equation}
\begin{array}{ll}
\displaystyle \label{amplitude} \partial_t g &= - 2 W_x \sin(2\partial_x \theta)\sinh(2\partial_x g)\\
\displaystyle \label{phase} \partial_t \theta &= - D_x + 2W_x \cos(2\partial_x \theta)\cosh(2\partial_x g)
 \end{array}
\end{equation}
The position of the peak $x(t)$ corresponds to the maximum of the amplitude, thus it satisfies the implicit equation $\partial_x g|_{x(t)} = 0$. To obtain its evolution, we differentiate it with respect to time:
\begin{equation}
\label{first}
\frac{dx(t)}{dt} \partial^2_x g + \partial_t \partial_x g = 0
\end{equation}
The term $\partial_t \partial_x g$ above can be computed from (\ref{amplitude}). At the point where $\partial_x g|_{x(t)} = 0$, (\ref{first}) becomes:
\begin{equation}
\label{evo_r}
\frac{dx(t)}{dt} = \left. 4 W_x \sin(2\partial_x \theta)\right|_{x(t)}
\end{equation}
This equation involves only one unknown quantity, $\partial_x \theta|_{x(t)}$, which, as we shall show, is akin to a momentum. A self-consistent equation for this quantity can be found taking its derivative with respect to time:
\begin{equation}
\label{second}
\frac{d\partial_x \theta|_{x(t)}}{dt} = \frac{dx(t)}{dt}\partial^2_x \theta|_{x(t)} + \partial_t \partial_x \theta|_{x(t)}
\end{equation}
In the previous expression, $\partial_t \partial_x \theta|_{x(t)}$ is the space derivative of (\ref{phase}). Using (\ref{evo_r}) the two terms in $\partial^2_x \theta|_{x(t)}$ cancel out, and (\ref{second}) reads:
\begin{equation}
\label{third}
\frac{d\partial_x \theta|_{x(t)}}{dt} = -\left. \partial_x D_x \right|_{x(t)} + 2 \cos(2\partial_x \theta|_{x(t)}) \left. \partial_x W_x \right|_{x(t)} 
\end{equation}
This, together with eq. (\ref{evo_r}), provides a set of closed differential equations for $x(t)$ and $\left. \partial_x \theta\right|_{x(t)}$.
The last thing we have to show is that the quantity $ \left. \partial_x \theta\right|_{x(t)}$ is the average value of the momentum operator $\hat{p}$, which reads:
\begin{equation}
\begin{array}{ll}
 \langle \hat{p} \rangle & \displaystyle  = - \int_x \frac{i}{V} \psi^\dagger_x\partial_x \psi_x \\
  & \displaystyle = i \int_x |\psi_x^2| \partial_x g(x,t) + \int_x |\psi_x^2| \partial_x \theta(x,t)\\
 \end{array}
\end{equation}
Because of the form (\ref{psiexp}) of $\psi_x(t)$ these integrals can be performed by the saddle point method (in the limit $V\rightarrow \infty$).
At the saddle point, $\partial_x g(x(t),t) = 0$ and $\partial_x \theta(x(t),t)$ has a non zero value, thus $\langle \hat{p} \rangle = \partial_x \theta|_{x(t)} + O(1/V)$. The same is true for $\langle \hat{x} \rangle = x(t) +  O(1/V)$. Thus,
rewriting (\ref{evo_r}) and (\ref{third}), one finds that the phase $p(t)$ and the position $x(t)$ obey the equations:
\begin{equation}
\begin{array}{ll}
\displaystyle \frac{dx(t)}{dt} &= 4 W_x \sin(p) = \partial_p (D_x-2W_x \cos(2p))\\
\displaystyle \frac{dp(t)}{dt} &= - \partial_x (D_x-2W_x \cos(2p))
 \end{array}
\end{equation}
As a conclusion, the average values $x(t)$ and $p(t)$ of the position and momentum operator, which describe the global behavior of the wave packet, have the same time evolution as classical canonical variables with Hamiltonian $H[x,p] = D_x - 2 W_x \cos(2p)$.
This property fits into the general picture of the propagation of wave packets in the semi-classical regime (see for example \cite{lit} for a mathematically oriented review) and, hence, was expected on general grounds as discussed previously. The generalization to more than one variable is straightforward. The width of the packet can also be evaluated. It is related to the separation in time of two classical trajectories initially separated by the width of the packet.
When the effective dynamics is one dimensional, the classical Hamiltonian is integrable and periodic orbits separate linearly in time.
In this case, since trajectories start from an initial distance $1/\sqrt{V}$, the typical time of separation is $t \sim \sqrt{V}$.
We check numerically that our analysis of the limit of large $V$ is correct. In figure \ref{fig:comparison}, the exact quantum evolution, obtained by diagonalization of the discrete equation \eref{gen_schrod}, is compared to the classical evolution for short times. We find an excellent agreement.

\subsection{Different effective trajectories in the phase space}
\label{branches}
We now study the effective classical dynamics by focusing on the phase space properties. \\
Since the Hamiltonian \eref{classical_Hamiltonian} is one-dimensional, and energy is conserved there are lots of constraints on the motion. In particular all the trajectories are integrable, and can only be periodic in $x(t)$ except for separatrix trajectories.\\
We find that the phase space is divided in two regions by a separatrix. To see this, one can characterize trajectories in terms of their turning points, and considering whether the momentum $p$ is bounded or not.
There are three types of turning points: the derivative $\frac{\rmd x}{\rmd t} = 4 W(x)\sin(2p) $ vanishes if either $p = 0$, $p = \pi/2$ or $W(2x') = 0$. This last case is called ``absorbing'' for a reason that will be explained later.

\begin{figure}
\centering
\includegraphics[scale=1.]{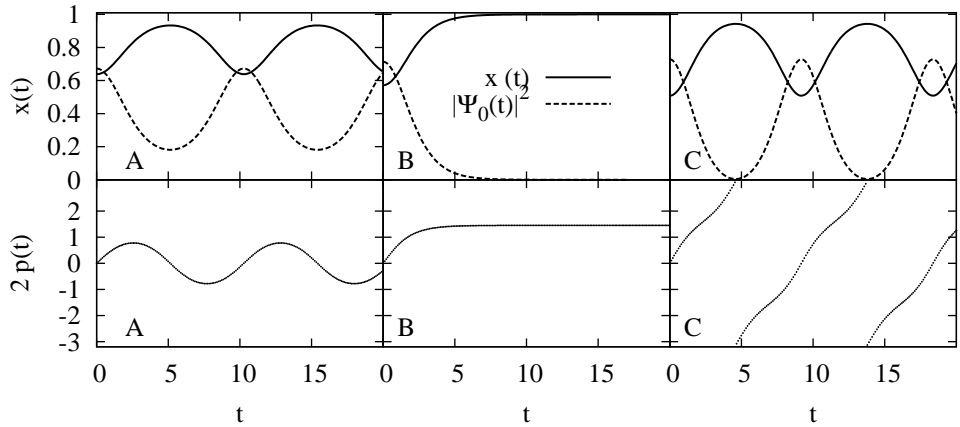}
\caption{The time evolution $x(t)$ and $2 p(t)$ modulo $\pi$ for the three trajectories $A$, $B$, $C$. The scale is the same for all graphs.}
\label{fig:traj_sample}
\end{figure}

\begin{figure}
\centering
\includegraphics[scale = 0.9, bb= 0 30 424 180]{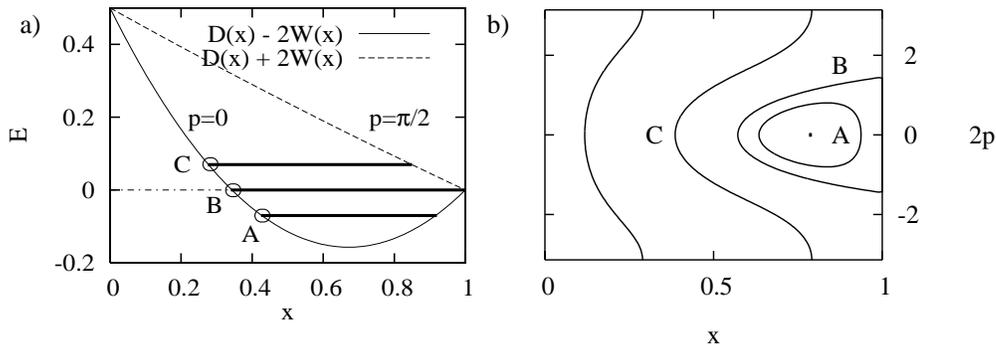}

\caption{a) Three trajectories $A$, $B$ and $C$ for $U = 3.33 < U_c$. The energy of each trajectory, $D+2W$ and $D-2W$ as functions of $x$. b) Trajectories in the phase space, momentum $2p$ versus $x$.}
\label{fig:traj}

\end{figure}

Let us compare three trajectories at different energies for $U<U_c$, which are representative of all the cases encountered.
For this purpose, two representations are shown: the evolution of $x(t)$, $p(t)$ in figure \ref{fig:traj_sample}, the phase space in figure \ref{fig:traj} (right panel).
The three types of trajectories are:
\begin{itemize}
\item (A) has two $p=0$ turning points, and thus its momentum $p$ is bounded.

\item (B) is a separatrix trajectory in the classical mechanics sense. The left turning point is at $p=0$, and the right turning point at $x=0$ is ``absorbing'', the time taken to reach it (or escape from it) is infinite. In the equation of the motion, at the point $W(x)=0$, the effective mass tend to infinity.

\item (C) has one turning point type at $p=0$ and one at $p=\pi/2$. $p$ is growing infinitely large with time, which is not pathological because only $p$ modulo $\pi$ enters the equations.
\end{itemize}
The diagram in figure \ref{fig:traj} (left panel) is similar in the spirit to the figures of effective potential shown to explain central motions in textbooks. It allows one to understand the dynamical evolution in a simple way. Actually, since the classical energy $E = D(x)-2W(x) \cos(2p)$ is conserved, the values of $x$ during the motion are restricted by the conditions $D(x)-2W(x) \leq E \leq D(x)+2W(x)$ (note that $W(x)\neq 0$ for $x\neq1,0$). 
Except for the separatrix, all turning points correspond to either $p=0$ at $E=D-2W$, or $p=\pi/2$ at $E=D+2W$. Thus, all trajectories are delimited by the two turning points $x_a$ and $x_b$, which are at the intersection between $E$ and $D(x) + 2W(x)$, and $E$ and $D(x) - 2W(x)$. The evolution of the system consists then in a periodic motion oscillating between $x_a$ and $x_b$.
\label{types}

\begin{figure}[htbf]
\centering
  \includegraphics[width = 12cm, bb=50 50 400 195]{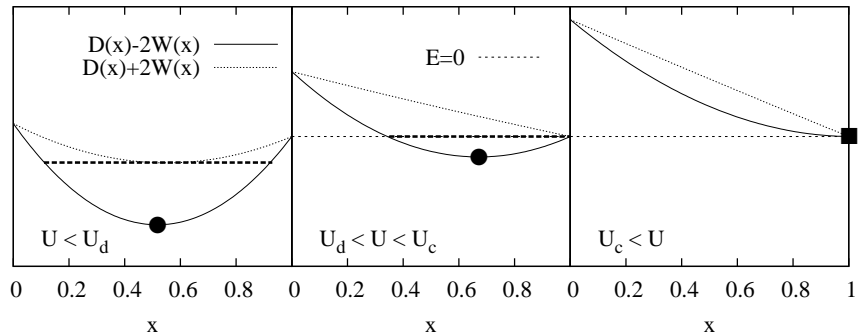}
\caption{$D(x) \pm 2W(x)$, versus $x$ for the three regimes of couplings $U$ in the Bose-Hubbard model $n_b^{max}\le 2$. Dark circles and squares are the superfluid and Mott insulator ground states respectively. Dark dashed lines correspond to separatrix trajectories. 
}
\label{fig:4}
\end{figure}

In the previous discussion we considered a specific form of the phase space and of  $D-2W$, $D+2W$ which are 
valid for certain values of $U$ only. In the next section a careful investigation of the $U$ dependence will be 
presented. However, before that, we want to stress that the effective potentials  $D-2W$, $D+2W$ can take three different qualitative form, depending on the coupling $U$. These are shown in figure \ref{fig:4}. The three regimes are separated by two special values of $U$: $U_d$ and $U_c$, see figure  \ref{fig:4}. 
$U_c$ is the critical coupling of the quantum phase transition, and $U_d = 1/U_c \simeq 0.1715$.
%
%
The special regime $U<U_d$ disappears when larger filling numbers $n_b^{max} \geq 3$ are included, so it is a peculiarity 
of the $n_b^{max}=2$ case and, hence, not very relevant. 

\section{Sudden quenches in the Bose-Hubbard model for $n_b^{max}=2$}
\label{quenches}

We now discuss the dynamical evolution following a quantum quench and its dependence on the final and initial
value of $U$. At $t<0$ the system is in the ground state at the coupling $U_i$.  At $t=0$ the coupling is switched to $U_f$ and the quench dynamics is computed for $t>0$.
In the following we call a quench ``from superfluid'' when $U_i < U_c$ and ``from Mott'' when $U_i > U_c$, and ``to superfluid'' or ``to Mott'' when $U_f >U_c$ and $U_f < U_c$. The results of the following sections are summarized by the dynamical phase diagram shown in \ref{fig:summary} (left panel).

\subsection{Mott to superfluid}
\label{Mott_to_superfluid}
Starting from the Mott ground state $x_0 = 1$, the trajectory is stuck at $x(t) \sim 1$ even for large times. In order to check this, let us linearize the equation of motion around $x=1$. 
We use that $\dot{x} = \partial H /\partial p =  4W_x \sin(2 p)$ and that $\sin(2 p)$ can be extracted from $E = D_x - 2W_x \cos(2 p)$ to obtain $\dot{x} = - \sqrt{4W^2_x - D_x^2}$. Thus, at dominant order in $\epsilon$ we obtain the equation for $\epsilon=1-x$:
\begin{equation}
\label{tau}
\dot{\epsilon} = \epsilon /\tau, \quad \tau =2/\sqrt{(U_c-U)(U-U_d)}
\end{equation}
 The trajectory $\epsilon(t)=0$ is unstable, and since the wave function has a width $1/\sqrt{V}$, its typical evolution is given by $\epsilon(t) =1/\sqrt{V} e^{t/\tau}$.
 Therefore, in the effective picture, the trajectory is stuck at the Mott ground state on times of the order of $\log(V)$.
This result is a peculiarity, actually a pathology, of mean field models. Indeed, in a real finite dimensional system, spatial fluctuations drive the system away from the Mott state in a finite time. 

\subsection{Superfluid to Mott}
In a superfluid to Mott quench, the initial condition is given by the ground state packet characterized by 
$\{ x = \frac{U/U_c +1}{2}, p = 0 \}$. The trajectory after the quench is of the type (C), see quench Q1 of figure \ref{fig:quenches}. The superfluid order parameter $|\Psi^2_0|$ (equation \eref{superfluid_order}) oscillates. 

\subsection{Superfluid to superfluid $U_d<U_f<U_c$}
\begin{figure}
\centering
  \includegraphics[width = 14cm, bb=64 57 400 195]{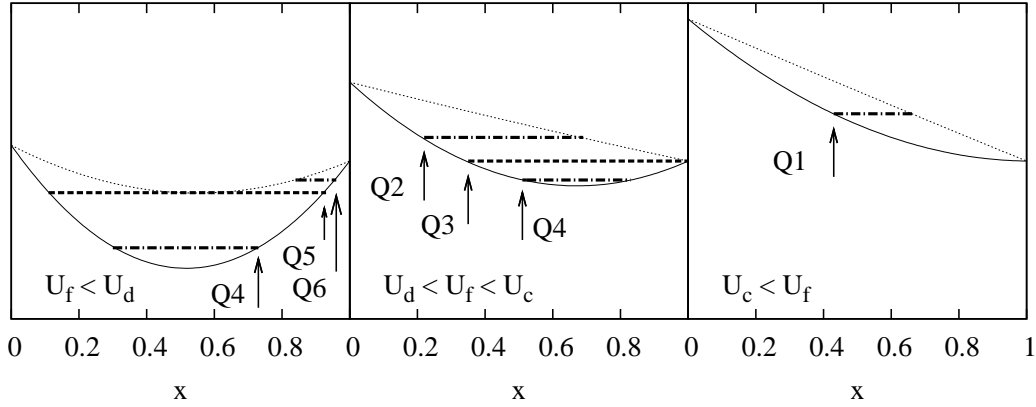}

\caption{Quench trajectories for the three regimes of $U_f$ with $D(x)\pm2W(x)$, like in figure \ref{fig:4}. Right panel ($U_c<U_f$):
In the quench Q1, at $t=0$ the packet state is at $x=x_0, p=0$ (thus on the line $D-2W (U_f)$), a position indicated by the arrow $Q1$.
Then the trajectory of constant energy $E$ is figured by the horizontal dark dash-dotted line. Left and center panel ($U_f<U_d$ and $U_d<U_f<U_c$): Different initial conditions lead to qualitatively different quenches. The dark dashed lines are singular trajectories of infinite period.
Quenches on these trajectories (Q3 and Q5) are at the dynamical transition.}
\label{fig:quenches}
\end{figure}

Depending on the value of $U_i$ and $U_f$, there are three different types of dynamical evolutions in this case, that we call Q2, Q3 and Q4 in figure \ref{fig:quenches}.  Q2 is of type (C) and Q4 is of type (A).
Q3, reached from a special $U^d_i$ (dependent on $U_f$), corresponds to a singular separatrix of type (B) and of energy $E=0$.
In this case the packet relaxes exponentially to the \emph{Mott insulator ground state} $|x=1 \rangle$, and concomitantly the superfluid order $|\Psi_0|^2$ vanishes exponentially in time.
This may be seen from the linearization \eref{tau}, which becomes $\dot{\epsilon} = -\epsilon /\tau$ for a trajectory relaxing to $x=1$. Approaching the transition 
oscillations take place on a time scale that diverges as $- \tau \ln(|U^d_f - U_f|)$.
In consequence we find that a dynamical singularity, or transition, takes place at $U^d_f$. The values of $U^d_f$
depends on $U_i$ and thus defines a dynamical transition line $U^d_f = (U_c + U^d_i)/2$, in the $U_i,U_f$ plane. 
In figure \ref{fig:summary} (right panel), as an example of singular behavior, we show the time average $\langle|\Psi_0|^2\rangle$ as a function of $U_f$ for quenches starting from the non interacting case $U_i=0$.\\
It is interesting to compare to $\langle|\Psi_0|^2\rangle$ its equilibrium counterpart obtained 
from the microcanonical average corresponding to the same energy. From this we clearly see
that the system is not thermalized. Actually, large quenches (Q2) have non vanishing oscillations of the superfluid order
contrary to the equilibrium value which is instead zero (because it corresponds to an effective high temperature). 
Moreover, we find that the dynamical transition takes place at an energy at which the system, if it were relaxed, would be superfluid.
Thus, the exponential relaxation of the superfluid order is a purely \emph{dynamical effect}. 
\subsection{Superfluid to superfluid $U_f<U_d$}
\begin{figure}
\centering
\includegraphics[scale=1.05, bb = 0 0 363 125 ]{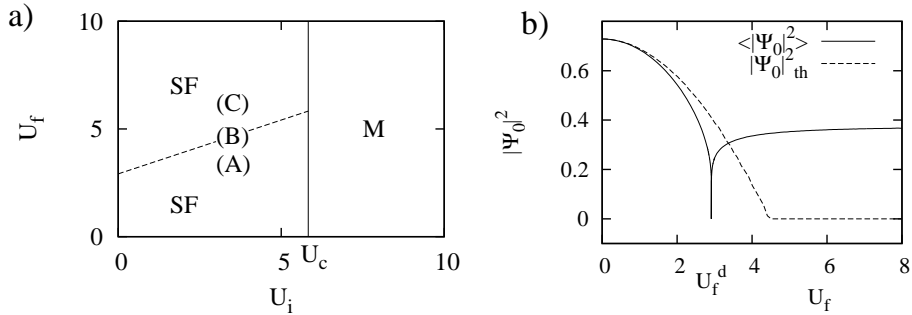}

\caption{a) Dynamical phase diagram for the Bose-Hubbard model with $n_b^{max}=2$. In the $M$ area, within mean field, the system remains stuck to the Mott insulator ground state after the quench. Quenches from the superfluid phase are oscillating and similar to (A) or (C).
The dynamical transition (B) separating the two is displayed as a dashed line, it meets the Mott phase at $U_f = U_c$.
b) Superfluid order $\langle |\Psi_0|^2\rangle$ as a function of $U_f$. Continuous line: time average after a quench. Dashed line: microcanonical average at the corresponding energy after the quench. The initial coupling is $U_i=0$, but the evolution is qualitatively similar for all $U_i$ with a dynamical transition.}
\label{fig:summary}
\end{figure}

In this regime, there is a qualitatively new family of quenches Q6 (comparable to Q2). There is also a new special quench Q5 to the modified separatrix state at energy $E_d$, which gives rise to another dynamical transition.
However, unlike the other quench cases, these effects are artifacts of the truncation $n_b^{max}=2$ 
and disappear for $n_b^{max} \geq 3$.
Therefore, they are not indicated in figure \ref{fig:summary} (left panel), where the outcome of all possible sudden quenches is summarized.

\section{Bose-Hubbard model with weaker truncation $n_b^{max} \geq 3$}
\label{more}
\subsection{Hamiltonian}
The previous analysis focused on two bosons or less per site. This constraint can be relaxed to include up to $n_b^{max}$ number of bosons per site. When $n_b^{max}$ is sufficiently high compared to the density, $n_b^{max} \gg n$, 
one recovers the behavior of the Bose-Hubbard model with no constraints on the occupation number per site. 

For a given maximum number of bosons per site, $n_b^{max}$, any symmetric wave function can be parametrized by the fractions $x_i$ of sites with $i$ bosons per site, $i \in [0,n_b^{max}]$. Since the $x_i$ are fractions, they verify $\sum_i x_i=1$. Moreover the density $n = \sum_i i x_i$ is fixed, thus there are only $n_b^{max}-1$ free variables left.
The wave function is expanded in the symmetric basis like $|\psi\rangle = \sum_x \psi_x(t) |x\rangle$. The transition elements $D(x)$ and $W_m(x)$ can be computed as done previously. For instance, there are 3 different types of transitions $m$ when $n_b^{max}=3$ and 6 when $n_b^{max}=4$.
Performing the classical equivalence for packet states, the resulting Hamiltonian can be put in the form \eref{classical_Hamiltonian_generic}:
\begin{equation}
H(x,p) = D(x) - 2 \sum_m W_m(x) \cos(m_i p_i)
\end{equation}
Specifically, when $n_b^{max} = 3$, if one chooses $x_1$ and $x_2$ as free variables, the Hamiltonian is
\begin{equation}
\hspace{-1.5cm}
\begin{array}{l}
H(x_1,x_2,p_1,p_2)= D - 2  W_1 \cos(p_1+p_2)- 2  W_2 \cos(2p_1-p_2)- 2  W_3 \cos(p_1-2p_2)\\
W_1 = J (3 x_0 x_1 x_2 x_3)^{1/2} \quad W_2 = J x_1(2 x_0 x_2 )^{1/2} \quad W_3 = J x_2(6 x_1 x_3 )^{1/2}\\
D = x_2 + 3 x_3 - J(x_0x_1+2x_1x_2+3x_2x_3)\\
x_0 = 1 - x_1 - x_2 - x_3 \qquad x_3 = \frac{1}{3}(n - x_1 -2x_2)
\end{array}
\end{equation}
Notice that because $0<x_i<1$, there are constraints on the possible values of $x_i$, such as $x_2<(n-x_1)/2$.
For all $n_b^{max}$, there is a Mott insulator to superfluid quantum phase transition at some coupling $U_c$ if the density $n$ is an integer (lower than $n_b^{max}$). Above the critical coupling $U>U_c$ the ground state is a Mott insulator $x_n = 1, x_{i \neq n} = 0$ (all sites have $n$ bosons), and below $U<U_c$ the ground state is superfluid with all $x_i \neq 0$.

\subsection{Regularity of trajectories after a quench}

\begin{figure}

\includegraphics[scale=0.87,bb = 1 35 486 185]{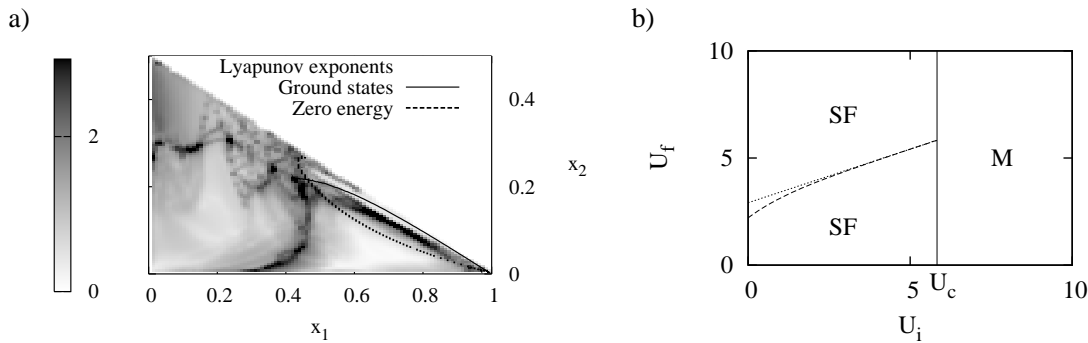}

\caption{a) Lyapunov exponents of the trajectories with initial conditions $\{x_1,x_2,p_1 =p_2 =0\}$, for $U=2.86$ plotted in levels of gray. The bright zone is regular (numerically $\lambda \lesssim 0.05$), the dark zones are chaotic.
The Lyapunov exponents corresponding to initial conditions given by ground states obtained varying $U_i$ are 
indicated by a continuous line. The dashed line indicates initial conditions with zero energy. The dynamical transition corresponds to their intersection.
b) Dynamical phase diagram for $n_b^{max} = 4$. The dynamical transition is at the dashed line. The dotted line is the dynamical transition when $n_b^{max}= 2$, for comparison.}
\label{fig:lyap}
\end{figure}

In the previous case $n_b^{max}=2$, the effective dynamics was one-dimensional, and thus integrable. For $n_b^{max}>2$, the classical dynamics takes place in two or more dimensions, and the trajectories may be either regular or chaotic.
In order to characterize them we study their regularity properties. For chaotic trajectories, neighboring trajectories separate exponentially in time in the phase space $y = \{x_i, p_i\}$, like $\delta y(t) \sim \exp(\lambda t)\delta y(0)$. The rate of separation $\lambda$ is the largest Lyapunov exponent \cite{lyapunov}.
We computed $\lambda$ for $n_b^{max}=3$ at density $n=1$ for several trajectories, using a simplified version of the traditional Gram-Schmidt orthonormalization of the Lyapunov vectors.
Roughly speaking, if we write the Hamiltonian evolution under the form $\dot{y_i} = f_i(y)$, the deviation satisfies $\dot{\delta y_i} = \partial_j f_i(y) \delta y_j$. This can be integrated numerically and normalized at each step to avoid an overflow (the orthogonalization step is dedicated to find all Lyapunov exponents, whereas here we only need the largest one).
The Lyapunov exponents for trajectories with different initial conditions $\{x^i_1,x^i_2,p_1 =p_2 =0\}$ are plotted in figure \ref{fig:lyap} in the superfluid phase $U=2.86<U_c$.
Trajectories are regular in some regions of the space (periodic or quasi periodic) with $\lambda = 0$ within the error bar, whereas some other regions are chaotic with $\lambda>0.1$.
The quench from $E=0$ is exactly at the dynamical transition, the corresponding trajectory is chaotic.
For $U > U_c$, when the ground state is a Mott insulator, all trajectories are regular.
We notice that the regularity of a trajectory affects the time of spreading of the packet (determined by the time of separation of two neighboring trajectories). Actually, the time of separation is typically polynomial $t \sim V^{1/\alpha}$ for a regular motion but only $t \sim \ln {V}$ for a chaotic one.
 
\subsection{Dynamical transition}

\begin{figure}
  \begin{center}
    \includegraphics[width=14.cm,bb = 52 52 435 224]{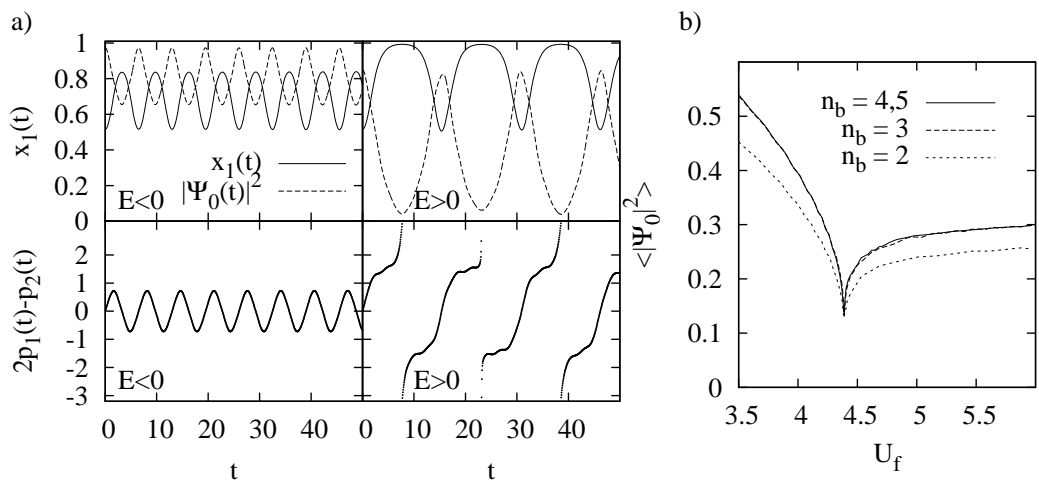}
   \end{center}
 \caption{\label{fig5}
 a) Evolution of $x(t)$ and $p(t)$ with time, $n_b^{max} = 3$, $n=1$ and $U_i=1$. Left panel, $U_f=2.5$, and  right panel, $U_f = 3.29$. The dynamical transition is at $U_f^d = 3.21$. The left panel is before ($E<0$) the dynamical transition, the right one ($E>0$) is after. b) Superfluid order parameter $\langle |\Psi_0|^2 \rangle$ as a function of $U_f$ for $n=1$, $U_i=3$, with $n_b^{max} = 2,3,4$ and $5$. The curve $n_b^{max} = 2$ is shifted of $0.025$ along the $U_f$ axis for comparison.}
 \end{figure}

In quenches from the superfluid phase $U_i<U_c$, like in the previous case $n_b^{max}=2$, a dynamical transition occurs at the special coupling $U_f^d$ where the final energy equals the energy $E=0$ of the unstable Mott trajectory $x_n = 1, x_{i \neq n} = 0$.
Direct evidence of this transition is given by the singularity of the time averaged superfluid order $|\Psi_0|^2$ as a function of $U_f$ for a given $U_i$.
In figure \ref{fig5}b, we compare this singularity in $|\Psi_0|^2$ for $n_b^{max} = \{2,3,4,5\}$ and find that the dependence in $n_b^{max}$ of the divergence is very weak beyond $n_b^{max} = 4$. The two curves $n_b^{max}=4$ and $n_b^{max}=5$ are not distinguishable, because they are identical up to $0.01\%$.\\
We show the dynamical phase diagram in figure \ref{fig:lyap}b for $n_b^{max}=4$ and unit filling factor. We observe that the transition line is near to the transition line for $n_b^{max}=2$, and that they are asymptotically equal around $U_c$.
Because the probability of having more than 4 bosons on the same site is extremely small, of the order of $0.01\%$ when $n_b^{max} \gg 4$, we can safely assume that this phase diagram is quantitatively representative of the phase diagram \emph{without truncation}.

Even though the existence of the dynamical transition for any $n_b^{max}$ is beyond doubt, because the singularity is numerically manifest, more precise results on this transition, even for $n_b^{max}=3$, are hard to provide.
Some features of the case $n_b^{max}=2$ persist ; for example, at the dynamical transition, the momentum $2p_1-p_2$ becomes unbounded, see Fig. \ref{fig5}a. The fractions $x_i(t)$ are also oscillating, but without definite period. They are either quasi-periodic for low $\lambda$ regions, or chaotic.
Approaching the dynamical transition, the typical time of return to $x_1 \sim 1$ is increasing, possibly logarithmically divergent in $U_f - U^d_f$ as suggested by the numerical divergence in figure \ref{fig5}b.
Unfortunately, the analysis of the singularity for $n_b^{max}\geq3$ turns out to be out of reach. The numerical integration of classical equations of motions is exponentially sensible to numerical errors and thus not reliable.
One could imagine that the trajectory at $E=0$ is an unstable manifold, which would support the existence of a singularity for trajectory arbitrarily close to it.
However, it is hard to decide whether the trajectory at the dynamical transition goes arbitrarily close to the point $x_1 = 1$ (ground state of the Mott insulator), at which the trajectory is exponentially slowed down.
A possible scenario, in which the surface $E=0$ is ergodic, does not seem to be validated by numerical integration of trajectories. 

\section{Quenches in the generalized Jaynes-Cummings model for super-radiance transition}
\label{Jaynes}

\subsection{Super-radiance quantum phase transition}

We now consider the generalized Jaynes-Cummings model \eref{Jaynes_Hamiltonian} and derive its quench dynamics. 
For simplicity and without loss of generality we focus on $\hat{Q} = N/2$. 
Thanks to the conservation of $\hat{Q} = \hat{S}^z + \hat{b}^{\dagger}\hat{b}$, the spin degrees of freedom can be accounted for in terms of the bosonic ones. 
The states can be parametrized by the density of bosons $n$ only, in the Fock basis $|n\rangle_{\indexit{bosons}} \otimes |S; m \rangle_{\indexit{spin}}$:
\begin{equation*}
|n \rangle = |Nn\rangle_{\indexit{bosons}} \otimes |N/2; N(1/2-n) \rangle_{\indexit{spin}}
\end{equation*}
The action on $|n\rangle$ of some useful operators can be readily computed at the dominant order in $N$
\[
\begin{array}{ll}
\hat{S}^z |n\rangle & = N(1/2-n)|n\rangle\\
 \hat{b}^{\dagger}\hat{b}|n\rangle & = Nn|n\rangle\\
\hat{S}^+ \hat{b} |n\rangle & =  \sqrt{nN}\sqrt{N^2/4 -m^2} |n-1/N\rangle\\
 & = N^{3/2}n\sqrt{1-n} |n-1/N\rangle
 \end{array}
\]
From these equations, the transition elements $D_n =\frac{1}{N} \langle n|\hat{H}|n\rangle$ and $W_n = - \frac{1}{N}\langle n\pm1/N|\hat{H}|n\rangle$ are easy to compute, and yield an effective Hamiltonian on $n$ and $\phi$ its conjugate momentum $H[n,\phi] = D_n - 2W_n \cos(\phi)$
\begin{equation}
\label{Jaynes_effective}
H[n,\phi] = -2 \lambda n +2 n\sqrt{1-n}\cos(\phi)
\end{equation}
which is the semi-classical approximation of the original Hamiltonian obtained in \cite{Gurarie}, up to a shift in the phase $\phi \rightarrow \phi+ \pi$. In \cite{Gurarie}, the authors studied sweeps from the empty state $\langle \hat{n} \rangle = 0$, and the link to the Landau-Zener problem. Here, we focus on sudden quenches and the related dynamical transition. The ground state of this effective system is at $\phi_{\indexit{GS}} = \pi$ and
\begin{equation*}
\hspace{-2cm}n_{\indexit{GS}} = \left \{
\begin{array}{lll}
\displaystyle 0 & \textrm{ if } \lambda \leq -1, & \textrm{Standard ground state}\\
\frac{2}{9} (3 - \lambda^2 - \sqrt{\lambda^2 (3 + \lambda^2)} & \textrm{ if } -1 \leq \lambda \leq 0 & \textrm{Super-radiant ground state}\\
\frac{2}{9} (3 - \lambda^2 + \sqrt{\lambda^2 (3 + \lambda^2)} & \textrm{ if } \lambda \geq 0 & \textrm{Super-radiant ground state}
\end{array} \right.
\end{equation*}

The super-radiance quantum phase transition is the result of a competition between the potential energy term and the kinetic energy term, which can be understood if the quantum Hamiltonian is written in the $|n\rangle$ basis:
$$
\hat{H} = -2\lambda N\hat n + \frac{g}{\sqrt{N}}N\sqrt{\hat{n}(1-\hat{n})}(\hat{b}^{\dagger} + \hat{b})
$$
Qualitatively, the potential term $-2\lambda \hat{b}^{\dagger}\hat{b}$ encourages filling for $\lambda \geq 0$ and discourages it for $\lambda \leq 0$. The ``kinetic'' contribution $\sim g(\hat{b}^{\dagger} +\hat{b})$ can lower the energy.
As a matter of fact, the sign of $g$ is almost irrelevant. Changing the sign of $g$
leads to $W_n \rightarrow -W_n$. However, the new Hamiltonian can be mapped into the old one by the translation $p \rightarrow p + \pi$.
In consequence, if the ground state of a given Hamiltonian is at $p = 0$, $\psi_n \sim \exp[-(n-n_{\indexit{GS}})/(\sigma/\sqrt{N})^2]$, then the ground state of the Hamiltonian with the opposite sign of $g$ is at $p=\pi$, $\psi_n \sim \exp[-i N \pi n-(n-n_{\indexit{GS}})/(\sigma/\sqrt{N})^2]$, identical but with a staggered sign.

\begin{figure}
\centering
  \includegraphics[width=14cm]{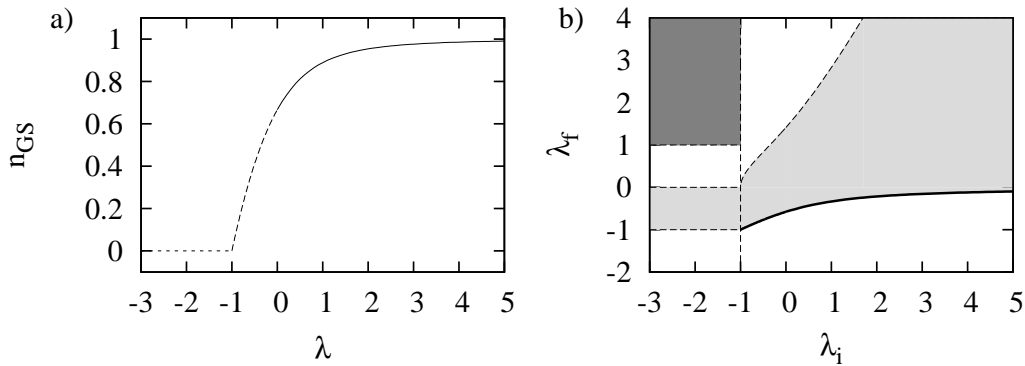}
\caption{a) Ground state $n_{\indexit{GS}}$ as a function of the parameter $\lambda$. The quantum phase transition takes place at $\lambda = -1$. b) Quench diagram for all $\lambda_i \rightarrow \lambda_f$. The quench trajectory is of bounded momentum $\phi$ (light gray) or unbounded $\phi$ (white). On the dashed line, the separatrix states are not singular (absorbing). The thick line is the dynamical transition, where the trajectories are singular. For $\lambda_i<-1$, in the dark gray region, trajectories are definitely stuck at $n=0$. Otherwise, there is a slow relaxation on times of order $\log N$.}
\label{fig:Jaynes}
\end{figure}

\subsection{Sudden quenches}
Previously, we have shown that the characterization of turning points plays an important role for understanding the dynamical behavior. In the generalized Jaynes-Cummings model, there are two special turning points, $n_c=\{0,1\}$, identified by the equation $\frac{dn}{dt}=W_n \sin(\phi)= 0$. Like in the Bose-Hubbard model, only one gives rise to singular trajectories. This is readily seen, for example, computing the time taken to reach the singular point and having zero kinetic energy at this point:
$$
T = \int^{n_c} \rmd n \frac{\rmd t}{\rmd n} = \int^{n_c} \rmd n (2 W_n \sin(\phi))^{-1}
$$
Using the conservation of the energy $E = D_n - 2W_n\cos(\phi)$, we can eliminate $\sin(\phi) = [1-(E(n_c)-D_n)^2/(4W_n^2)]^{1/2}$ and obtain
$$
T = \int^{n_c} \rmd n [4W_n^2 - (E(n_c)-D_n)^2]^{-1/2}
$$
Around $n = 1 -\epsilon$, the time is finite $T \sim \int_0 \rmd \epsilon \; 1/\sqrt{\epsilon}$, whereas around $n = \epsilon$, the time $T \sim \int_0 \rmd \epsilon \; 1/\sqrt{\epsilon^2}$ diverges. In other words, the trajectories touching the turning point $n = 1$ have a finite period, whereas the trajectories touching $n=0$ have infinite period. The situation is anologous to the one already studied for the BHM: the absorbing state $n_c= 0$ plays the same role of the Mott state, and gives rise to the dynamical transition.\\
To identify the critical value $\lambda_f^d$ at which the transition takes place, one can compute the energy after a quench $\lambda_i \rightarrow \lambda_f$ and compare it to the energy of the system at rest in $n=0$. One can check that a trajectory $n(t)$ starting with the same energy, actually energy zero, is indeed driven to the absorbing point $n_c = 0$. This dynamical transition occurs on the line $\lambda_f^d = -\sqrt{1 - n_{\indexit{GS}} (\lambda_i)}$. A linearization similar to \eref{tau} shows that the relaxation is exponential at the transition with a characteristic time $\tau^{-1} = 2\sqrt{1-\lambda^2}$. Around the transition, the period of a trajectory diverges as $ \tau \ln(|\lambda_f - \lambda_f^d|)$.

The dynamical phase diagram is shown in figure \ref{fig:Jaynes}b. For $\lambda_i>\lambda_c$, the initial state is in the broken symmetry phase $\langle \hat{n}(t=0) \rangle \neq 0$. The situation is very similar to the BHM one: the filling number $n(t)$ oscillates in time after the quench, and there is one region with bounded effective momentum $\phi$, and two regions with unbounded $\phi$. Notice that there are two separatrices between the three regions, and that one is \emph{not singular}, and does not give rise to a dynamical transition, whereas the other is. The singular one is the one for which the absorbing state $n=0$ is met, corresponding to the restored symmetry state. For quenches with $\lambda_i<\lambda_c$, the initial state is empty $\langle \hat{n}(t=0) \rangle = 0$, and either drifts away from zero on large times (of the order of $\log N$) if $\lambda_f<1$, and otherwise, the state is definitely stuck at $n=0$. The bounded or unbounded nature of the final state is also indicated.

Finally, we remark that eigenstates of the completely connected model can be written within a WKB expansion, this is done 
for completeness in \ref{wkb_expansion}. 

\section{Quenches in the Ising model in a transverse field}
\label{Ising}
In our last example, let us consider the dynamics due to quantum quenches in the transverse field Ising model. To derive the classical effective dynamics, one can use the fact that the large spin limit is also the classical limit, see e.g. \cite{mds}. To do so, one should make the substitution $\vec{\hat{S}}\rightarrow \vec{S} = S\{\sin\theta \cos\phi, \sin\theta \sin\phi, \cos\theta \}$. The kinetic term (needed to describe classical dynamics) in the classical Hamiltonian of the large spin limit can be derived using path integral for spins \cite{fradkinbook}.
Here we instead use our generic method based on site permutation symmetry. We denote spin symmetric states  of total momentum $S=N/2$ with the notation $|s\rangle$. They are such that $\hat{S}^z |s\rangle = Ns |s\rangle $ with $s \in [-1/2,1/2]$. At dominant order in $N$, $\hat{S}^x|s\rangle = N \sqrt{\frac{1}{4}-s^2}\left(|s+\frac{1}{N}\rangle +|s-\frac{1}{N}\rangle \right)/2$. One can proceed as usual to get the effective Hamiltonian:
$$
\begin{array}{ll}
D_s = \displaystyle \frac{1}{N} \langle s | \hat{H}|s\rangle & = -\displaystyle\frac{1}{2} s^2 \\
W_s = \displaystyle-\frac{1}{N} \langle s +\frac{1}{N}| \hat{H}|s\rangle & = \displaystyle\frac{\Gamma}{2}\sqrt{\frac{1}{4}-s^2}\\
 H[s,\phi] = D_s - 2 W_s \cos\phi &= -\displaystyle\frac{1}{2} s^2 -\Gamma\sqrt{\frac{1}{4}-s^2} \cos \phi\\
\end{array}
$$
As expected, the Hamiltonian is very reminiscent of the classical Hamiltonian for a single rotor once the change of variables $s \rightarrow \cos \theta$ is made. Indeed, the equations of motion obtained here are the same as is \cite{mds}. In this paper, the authors studied the quench from the paramagnetic phase to the ferromagnetic phase and AC dynamics.

\begin{figure}
\centering
  \includegraphics[width=14cm]{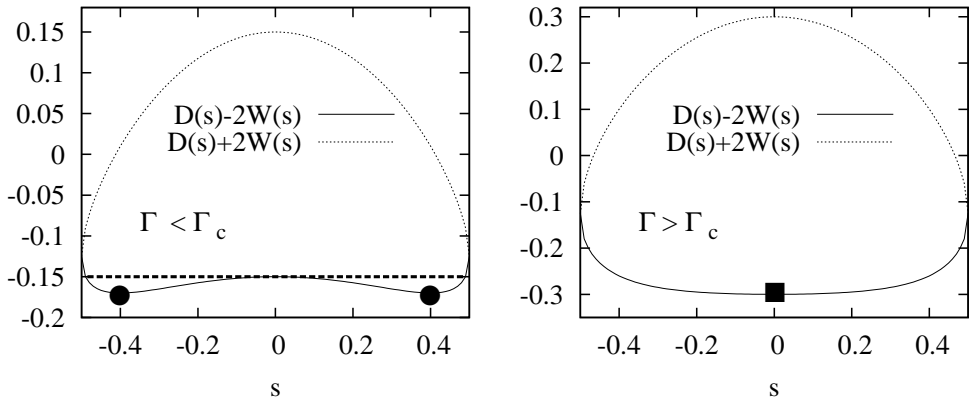}
\caption{$D(s)\pm 2W(s)$ for $\Gamma = 0.3 < \Gamma_c$ and $\Gamma = 0.6 > \Gamma_c$. The dark square is the quantum paramagnetic ground state, the dark circles are the ferromagnetic ground states. The dark dashed line is the separatrix, quenches on this trajectory are at the dynamical transition.}
\label{fig:ising}
\end{figure}

Proceeding as for the BHM, we analyze the phase space of symmetric trajectories. This is different in the ferromagnetic and paramagnetic phases, see figure \ref{fig:ising}. In the latter, obtained for $\Gamma < \Gamma_c$,  there is a separatrix.
For sudden quenches from the ferromagnetic phase, where $s \neq 0$, to final values of the transverse field $\Gamma_f < \Gamma_c$, there is a dynamical transition at $\Gamma^d_f = \frac{1}{4} + \Gamma_i^2$. At the transition, the trajectory is a separatrix, and $s$ decreases exponentially in time to $s=0$, which is the quantum paramagnetic ground state. The symmetry is dynamically restored at this point but also for larger quenches: contrary to the two previous examples, for quenches beyond the dynamical transition $\Gamma_f > \Gamma_f^d$, the trajectories are symmetric in $s \rightarrow -s$, and the magnetization is oscillating around zero. Averaging over time, the order parameter is zero $\overline{\langle \hat{S}^z \rangle} = 0$.

\section{Discussion}
\label{discussion}

Before concluding, two points need to be addressed further: (1) the possible link between dynamical transitions and  equilibrium quantum phase transitions and (2) the pro and cons of our approach concerning the physics of 
finite dimensional systems.

\begin{itemize}
\item In three examples, we found that dynamical transitions occur in systems characterized by a quantum equilibrium phase transition. Therefore, despite the fact that the dynamical transition is not directly related to the equilibrium one, as we stressed before, it is natural to wonder whether instead its {\it existence} is related to it. In support of this, if one studies the completely connected Bose-Hubbard model in a parameter regime (non unitary fillings) where there is no {\it equilibrium} quantum phase transition then one finds that there is no dynamical transition either. Actually, we found the very same phenomenon in all models we studied, and this was also noticed for the Fermionic Hubbard model \cite{schiro}. We certainly cannot address this issue in general, also because it is still far from clear what ``the dynamical transition'' is beyond mean-field theory. We shall instead endeavor to provide a partial answer at the mean-field level. 
Generically, we observe that in all models we studied, there is a regime in which 
the symmetric state (the Mott insulator, the paramagnet, etc.) is not the ground state but it
plays the role of an excited metastable state that can trap the system
for an infinite time. This gives rise to the dynamical transition. 
 If the effective motion is one dimensional, this state is absorbing and belongs to a separatrix. If the effective motion has two or more degrees of freedom, this state seems to belong to an unstable manifold. Thus, it appears that the singularity at the dynamical transition is a result of this singular trajectory, which in turn is a consequence of the existence of a quantum phase transition. A physical argument to support this idea is the following. For a quench from the unbroken symmetry phase to the broken symmetry phase in completely connected model, the dynamical evolution is expected to be very slow. The reason is that the symmetry must break up and the order must develop starting from this point. This, however, is a local phenomenon, for example due to domain growth in finite dimensions, and therefore generically absent in  completely connected models, except on times diverging with the system size: indeed, in our examples, we find that the typical time scale to depart from the initial value $x(t=0)$ is of the order of $\log{V}$. The point is that this slow trajectory is precisely the \emph{time reversal} of the singular trajectory that is responsible for the dynamical transition. To summarize, the existence of a quantum phase transition implies the existence of slow trajectories. They correspond to the escape from the symmetry unbroken state. Their time reversed counterparts are the trajectories responsible for the dynamical transition. {\bf Thus, the existence of a phase transition indeed appears to be 
intertwined with the existence of a dynamical transition within mean-field.} 

\item Let us now discuss the range of applicability of our results for finite-dimensional systems. First, our analysis can be put on the same footing as the Gutzwiller Ansatz and functional integral saddle point approximations, since the corresponding dynamical equations of motion coincide (see \ref{Gutzwiller} for the former equivalence and also \cite{snoek}).
Clearly, our mean-field approximation misses several essential physical effects. Relaxation and thermalization, for which spatial and temporal fluctuations must be taken into account, are not present. Furthermore, the dynamics of a quench from the unbroken symmetry phase to the broken symmetry phase is not properly described for the reasons mentioned above. Inhomogeneities, topological defects, domain growth are out of reach.

On the contrary, our mean field approximation is expected to capture well the evolution of local quantities at least on short times, as long as they are homogeneous across the system. Contrary to many other approaches, it is \emph{not perturbative} in parameters ($U$, $J$, $\Gamma$ \dots) of the Hamiltonian. {\bf Unlike pertubative approaches, it can describe the existence of dynamical transitions}, and phenomena related to global observables.
\end{itemize}

\section{Summary and outlook}
\label{conclusion}

The general purpose of this article was twofold. First, we described a method to study the quantum quench dynamics of generic completely connected models, providing a mapping to effective classical dynamics. 
There are two reasons for considering completely connected models. In some cases,  such as in the generalized Jaynes-Cummings model, and the Dicke model, they provide the correct physical description. In others, such as 
the transverse field Ising model or the Bose-Hubbard model, they lead to an approximative treatment of finite dimensional systems. In the latter case, the range of validity of the approximation is possibly limited to short times only.
Our second aim was to study and to reveal the existence of out of equilibrium dynamical transitions induced by quantum quenches. In agreement with other works \cite{schiro,calabrese} we showed that within the mean-field approximation dynamical transitions occur in quenches from the broken symmetry phase to other regions of the broken symmetry phase and that this is a quite generic phenomenon for systems that display a quantum phase transition at equilibrium. 

Clearly, a main and pressing question is to understand what this transition is really, beyond mean-field theory. 
Does it become a cross-over in finite dimensions? If it remains a bona fide transition, what are its critical properties? 
In the $d=1$ BHM, a non-monotonic behavior of the propagation velocity as a function of $U_f$ has been found in \cite{corinna2} by t-DMRG. This could well be a signature of a cross-over related to the dynamical transition found in mean-field.  
An exact diagonalization study of hard core bosons in $d=2$ suggests the existence of a cross-over or a singularity
of the revival time after a quantum quench \cite{wolf}. From the analytic point of view going beyond mean-field is a difficult task.  An argument for the occurrence of dynamical transitions beyond mean field is given in \cite{calabrese}, within an imaginary time path integral formalism. Clearly, it is crucial to take into account fluctuations not captured by mean-field theory. 
This was started to be done in \cite{schiro2}. This work indeed suggests that these fluctuations could alter substantially 
the mean-field results.  Promising ways to capture fluctuations are the projector operator formalism \cite{sengupta}, $1/z$ expansions \cite{Schutz} ($z$ is the connectivity of the lattice). Field-theoretic methods are also being developed, as in \cite{kennett} for the Bose-Hubbard model in the grand canonical ensemble; the use of two particle irreducible actions also seems promising \cite{calzetta}.


\ack
We thank C. Kollath, M. Schir\`o, M. Eckstein and J. Keeling for interesting discussions on these topics. G. Biroli acknowledges partial financial support from ANR FAMOUS.


\appendix
\setcounter{section}{0}

\section{Gutzwiller Ansatz}
\label{Gutzwiller}
The previous analysis, based on the symmetry between sites, seems to be related to completely connected models only. It is instead much more robust than it looks, since it is actually equivalent to the time dependent Gutzwiller Ansatz. The latter can be formulated as a mean field approximation for a finite dimensional model: it consists in 
providing a trial wave function dependent on some physical parameters that are chosen through a maximization procedure. In its equilibrium version, the estimated ground state is simply found minimizing the energy. The dynamical formulation is a bit more elaborated. For clarity, we show here how to build it for the case $n_b^{max}=2$ only. The following is an adapted version of the formulation developed in \cite{schiro}.

Let us define first the ``full state'' $|F \rangle = |0\rangle + |1\rangle +|2\rangle$, and let $\hat{P}^l_j$ be the projector on site $l$ on the state $|j\rangle$, where $j \in \{0,1,2\}$. The Gutzwiller Ansatz wave function is a trial wave function with $6$ real, time dependent parameters $x_j(t)$, $p_j(t)$:
\begin{equation}
\label{Gutzwiller_wave}
\begin{array}{ll}
| \psi_{\indexit{GA}} \rangle &= \displaystyle \prod_{\indexit{sites}} e^{i (p_0\hat{P}_0  + p_1 \hat{P}_1+p_2\hat{P}_2)}  \left( \sqrt{x_0} \hat{P}_0+ \sqrt{x_1} \hat{P}_1 + \sqrt{x_2} \hat{P}_2 \right) |F \rangle\\
&= \displaystyle \prod_{\indexit{sites}} \left(  e^{i p_0}\sqrt{x_0}|0\rangle +e^{i p_1}\sqrt{x_1}|1\rangle +e^{i p_2}\sqrt{x_2}|2\rangle \right) \\
\end{array}
\end{equation}
Note that for simplicity we dropped the site index. The $x_n$ are the projector average: $x_{n} = \langle \psi | \hat{P}_{n}|\psi \rangle$. As we shall see, they will turn out to coincide with the fraction of sites with $n$ bosons. Obviously, because this is only a trial wave function, it can not satisfy exactly the time evolution: i.e. $i \hbar \frac{\rmd }{\rmd t} | \psi_{\indexit{GA}}  \rangle \neq \hat{H} | \psi_{\indexit{GA}}  \rangle$. However, one can enforce it in an approximate way. This allows one to obtain the time evolution of $x_j(t)$ and $p_j(t)$. We impose the following constraints:
\begin{itemize}
\item The projectors in the Heisenberg representation $\hat{P}^H_l$ satisfy on average the Heisenberg equation of evolution $\dot{\hat{P}^H_l} = i [\hat{H},\hat{P}^H_l]$. This amounts to computing the average of $\hat{P}$ in the Schr\"odinger picture $x_l(t) = \langle \psi |\hat{P}_l| \psi \rangle$ using
\begin{equation}
\label{evolution_x}
\begin{array}{ll}
\dot{x_l} & = \dot{ \langle \psi |} \hat{P}_l | \psi \rangle +\langle \psi | \hat{P}_l \dot{ | \psi \rangle}\\
& = -i \langle \psi | [\hat{P}_l,\hat{H}]| \psi \rangle\\
\end{array}
\end{equation}
\item The energy is conserved $E = \langle \psi | \hat{H} | \psi \rangle$.
\end{itemize}
Let us call $\mathcal{H}[x_l,p_l] =  \langle \psi | \hat{H} | \psi \rangle$ the energy as a function of the parameters. The evolution \eref{evolution_x} can also be written, using the explicit form of $| \psi \rangle$ \eref{Gutzwiller_wave}:
\begin{equation}
\begin{array}{ll}
\dot{x_l} & = -i \langle \psi | [\hat{P}_l,\hat{H}]| \psi \rangle = \displaystyle \frac{\partial}{\partial p_l}\langle \psi | \hat{H} | \psi \rangle \\
 &= \displaystyle \frac{\partial \mathcal{H}[x_l,p_l]}{\partial p_l} \\
\end{array}
\end{equation}
This evolution is very reminiscent of the first Hamilton equation of motion. Furthermore, the conservation of energy $\dot{E} = \frac{\partial \mathcal{H}[x_l,p_l]}{\partial p_l} \dot{p_l} + \frac{\partial \mathcal{H}[x_l,p_l]}{\partial x_l} \dot{x_l}=0$ leads to the second Hamilton equation:
 \begin{equation}\dot{p_l} = -\frac{\partial \mathcal{H}[x_l,p_l]}{\partial x_l}\end{equation}

As a consequence, the Gutzwiller Ansatz wave function is indeed equivalent to an effective classical Hamiltonian 
evolution of the variables $x_l$ and $p_l$. The effective Hamiltonian $\mathcal{H}[x_l,p_l] = \langle \psi | \hat{H} | \psi \rangle [x_l,p_l]$ must be computed for the model at hand, and is indeed the same as the one discussed in section \ref{general}. For instance, for the Bose-Hubbard model with truncation $n_b^{max}=2$, one recovers \eref{classical_Hamiltonian}.

\section{Eigenstates in completely connected models}
\label{wkb_expansion}

Using a Wentzel-Kramers-Brillouin (WKB) approximation \cite{wkb}, eigenstates of the Schr\"odinger equation \eref{Schrodinger} can be found in the limit of large $V$. To do so, we write the eigenstates $\phi_E(x)$ within the WKB approximation. The derivation is standard, one looks for stationary solutions of the form $\psi(x) = A(x) e^{i V S(x)}$ of the equation \eref{Schrodinger}, at dominant order in $V$. One finds
%

\begin{equation}
\phi_E(x) = \frac{C}{2 W(x)\sqrt{\sin (2p(x))}} \exp \left( \pm i V \int^{x}_{x_0} \rmd x' p(x')\right)
\end{equation}
where $C$ is a normalization constant, $E$ is the energy, $p(x)$ satisfies the implicit equation $E = D_x- 2 W_x \cos(2p(x))$, and $x_0$ is the left turning point of the classical trajectory of energy $E$.

Using this expression, we can draw a parallel between the eigenstate $\phi_E(x)$ and the effective classical trajectories of energy $E$. For an observable $f(\hat{x})$, the average reads

\begin{equation}
\begin{array}{ll}
\langle \phi_E |f(\hat{x})|\phi_E \rangle &= |C|^2 \displaystyle \int_x \rmd x \frac{f(x)}{4 W(x)\sin (2p(x))} \\
& =  \displaystyle  \frac{1}{\int \rmd t} \;\int_x \rmd x \frac{\rmd t}{\rmd x} f(x) = \overline{f(x)}\\
\end{array}
\end{equation}
In the second line, we refer to the effective classical motion $\frac{\rmd x}{\rmd t} = \frac{\partial H}{\partial p}$, and the average $\overline{f(x)}$ is the average over the effective classical trajectory of $f(x)$ over one peridod.
In words, we found that in the semi classical regime, the average over a quantum stationary state of energy $E$ is given by the average over one period of the classical trajectory of same energy $E$. Although this result is not often mentioned, this is a direct consequence of the WKB expression of the wave function.

\end{document}